\documentclass{article}
\usepackage{ijcai26}

\usepackage{times}
\usepackage{soul}
\usepackage{url}
\usepackage[hidelinks]{hyperref}
\usepackage[utf8]{inputenc}
\usepackage[small]{caption}
\usepackage{graphicx}
\usepackage{amsmath}
\usepackage{amsthm}
\usepackage{booktabs}
\usepackage{algorithm}
\usepackage{algorithmic}
\usepackage[switch]{lineno}
\usepackage{xcolor}
\usepackage{lipsum}
\usepackage{dsfont}

\newcommand{\NAME}{\textbf{TEMG-TTA}}

\usepackage{latexsym}

\usepackage{amsfonts}
\usepackage{multirow}
\usepackage{colortbl}
\usepackage{placeins}
\usepackage{makecell}

\title{Temporal Motif-aware Graph Test-time Adaptation for\\ OOD Blockchain Anomaly Detection}

\author{
Runang He$^1$\and
Tongya Zheng$^{1,2,3}$\and
Huiling Peng$^1$\and
Yuanyu Wan$^{1,3}$\and
Bingde Hu$^{1}$\thanks{Corresponding Author}\and \\
Jiawei Chen$^{1,3}$\and
Canghong Jin$^{2}$\and
Mingli Song$^{1,2,3}$\and
Can Wang$^{1,2,3}$\and
\\
\affiliations
$^1$State Key Laboratory of Blockchain and Data Security, Zhejiang University\\
$^2$Zhejiang Provincial Engineering Research Center for Real-Time SmartTech in Urban Security Governance,  Hangzhou City University\\
$^3$Hangzhou High-Tech Zone (Binjiang) Institute of Blockchain and Data Security\\
\emails
herunang@zju.edu.cn,
\{doujiang\_zheng,phlnku\}@163.com,
\{wanyy,tonyhu,sleepyhunt\}@zju.edu.cn,
jinch@hzcu.edu.cn,
\{brooksong,wcan\}@zju.edu.cn
}

\begin{document}

\maketitle

\begin{abstract}
    Ever-evolving transaction patterns have significantly hindered anomaly detection on emerging cryptocurrency blockchains due to the vast number of addresses and diverse anomalous behaviors. 
    Recently, advanced Graph Anomaly Detection (GAD) approaches applied to blockchains have faced two critical challenges: \textit{adversarial pattern evolution by malicious actors} and \textit{the out-of-distribution (OOD) problem caused by varied transaction semantics on blockchains}.
    To address these challenges, we propose a novel framework termed \textbf{TE}mporal \textbf{M}otif-aware \textbf{G}raph \textbf{T}est-\textbf{T}ime \textbf{A}daptation (\NAME).
    First, we comprehensively capture the 3-node temporal motif distribution of each active address using an efficient computational mechanism, enabling downstream temporal motif-aware graph learning. Second, we design a simple yet effective test-time adaptation strategy to facilitate the sharing of common patterns between training and testing graphs.
    Extensive experiments on 5 real-world datasets demonstrate that our proposed \NAME~ outperforms \textit{state-of-the-art} GAD approaches by an average of 54.88\%.
    A further case study on interpretable motif patterns reveals that \NAME~ explicitly characterizes the complex transaction patterns of anomalous addresses, thereby verifying the effectiveness of our technical designs.
    Our code is publicly available at \url{https://github.com/LuoXishuang0712/TEMG-TTA/}.
\end{abstract}

\section{Introduction}

\begin{figure}
    \centering
    \includegraphics[width=0.45\textwidth]{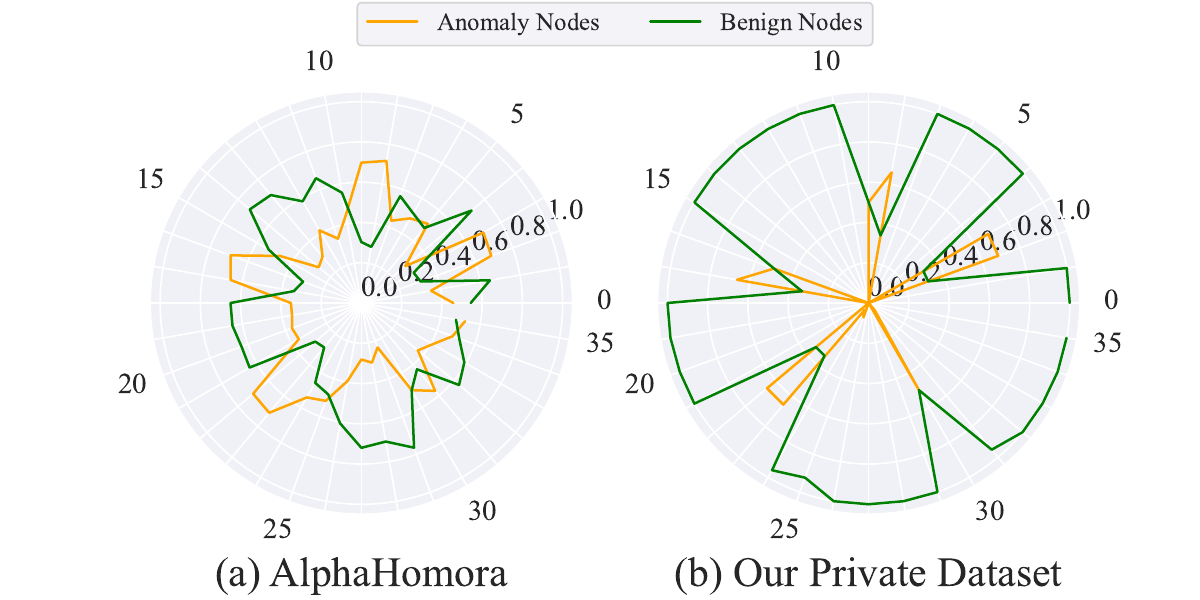}
    \caption{Radar maps of two types of blockchain transactions \textit{w.r.t.} 36 types of 3-node temporal motifs.}
    \label{fig:motifs_distribution}
\end{figure}

Blockchain transaction networks consist of large-scale, directed, and temporally ordered interactions, where anomalous behaviors continuously evolve. 
This dynamic nature makes blockchain anomaly detection particularly challenging under distribution shifts and limited supervision.
Cryptocurrencies on decentralized blockchains, such as Bitcoin and Ethereum, are widely used and support millions of transactions daily\footnote{\url{https://etherscan.io/txs}}. 
Although blockchain transparency enables public verification, it also allows illicit activities to hide within legitimate transaction flows, especially in programmable environments enabled by smart contracts~\cite{chen2020traveling}.

From a modeling perspective, blockchain transactions can be represented as directed temporal multi-graphs, motivating graph-based Blockchain Anomaly Detection (BAD) methods for asset flow analysis and anomalous group discovery~\cite{qi2023blockchain,luo2024crypto}. 
However, anomalous behaviors on blockchains are non-stationary and evolve with emerging cryptocurrency techniques and attack strategies~\cite{yao2024pulling}. 
Thus, models trained on historical data often suffer from severe out-of-distribution (OOD) issues in real-world deployment.

Early detection efforts relied on hand-crafted features and traditional machine learning models to detect Ponzi schemes~\cite{chen2018detecting}, trace cryptocurrency scams~\cite{phillips2020tracing}, and identify phishing transactions~\cite{yuan2020phishing}. 
With the development of deep graph learning, recent studies have shifted toward end-to-end representation learning for money laundering detection~\cite{weber2019anti} and account de-anonymization on Ethereum~\cite{zhou2022behavior}. 
More generally, Graph Anomaly Detection (GAD) methods have addressed challenges such as heterophily and label imbalance, as exemplified by DGAGNN~\cite{duan2024dga} and SpaceGNN~\cite{dong2025spacegnn}. 
Despite these advances, most existing methods rely on supervised learning and remain sensitive to distribution shifts in blockchain data.

Our closer inspection of transaction behaviors suggests that many complex patterns can be decomposed into fine-grained temporal motifs.
As shown in Figure~\ref{fig:motifs_distribution}, anomalous and benign nodes exhibit distinct temporal motif patterns on both the AlphaHomora dataset and a real-world money-laundering dataset.
Moreover, we find that improved detection performance is often accompanied by adaptive changes in adversarial behaviors, further exacerbating distribution shifts.
These observations reveal two unresolved challenges:
(i) existing GAD methods rarely model temporal motifs explicitly, limiting their ability to capture evolving adversarial patterns; 
(ii) supervised learning approaches struggle to adapt to structural changes under OOD settings.

In this work, we collaborate with Zhejiang Provincial Public Security Department, which provides reliable on-chain anomaly labels and cases, to detect OOD anomalies on blockchains using public transaction records and provided labels.
Specifically, we propose a \textbf{TE}mporal \textbf{M}otif-aware \textbf{G}raph \textbf{T}est-\textbf{T}ime \textbf{A}daptation (\NAME) framework to enable explicit temporal motif perception for OOD blockchain anomaly detection.
First, we design an efficient motif matching algorithm that reduces the time complexity from $\mathcal{O}(M^3)$ to $\mathcal{O}(M\cdot k^2)$, where $M$ is the number of transactions and $k$ is the maximum number of edges within a constrained time window.
Next, we construct a comprehensive temporal motif representation for each active node by incorporating shared motif prototype embeddings, role representations, and positional encodings.
Finally, we introduce a trustable node selection mask and a teacher-student regularization mechanism to facilitate graph TTA on blockchains and mitigate disruptive deviation.
Extensive experiments on four public datasets and one private dataset demonstrate that \NAME~ significantly outperforms \textit{state-of-the-art} GAD approaches by an average of 54.88\%.

Our main contributions are summarized as follows:
\begin{itemize}
\item We collaborate with a Provincial Public Security Department to detect suspicious Ethereum transactions in real time and report anomalous addresses with specified patterns, such as anomalous exchangers.
\item We propose \NAME, a temporal motif-aware graph test-time adaptation framework that improves structural expressiveness and robustness to temporal distribution shifts for blockchain anomaly detection.
\item Extensive experiments on multiple real-world blockchain datasets show that our method consistently outperforms state-of-the-art baselines and remains robust under temporal distribution drift. In-depth motif analysis and ablation studies further demonstrate the effectiveness of different components.
\end{itemize}

\section{Related Works}

\paragraph{Graph Anomaly Detection.}
Graph neural networks (GNNs) have shown strong effectiveness in graph anomaly detection (GAD)~\cite{qiao2025deep}. 
Recent studies address heterophily and imbalance in GAD, such as H$_2$GCN~\cite{zhu2020beyond}, PMP~\cite{zhuo2024partitioning}, and ConsisGAD~\cite{chen2024consistency}. 
Graph foundation models (GFMs), including UNPrompt~\cite{niu2024zero}, ARC~\cite{liu2024arc}, and AnomalyGFM~\cite{qiao2025anomalygfm}, further aim to improve generalization to complex abnormal patterns. 
However, these methods generally assume stationary distributions and struggle with continuously evolving blockchain transaction patterns.

\paragraph{Dynamic and Spatiotemporal Graph Modeling.}
Dynamic and spatiotemporal graph modeling has been widely studied for evolving behavioral patterns. 
Representative studies model temporal aggregation and propagation for dynamic graph representation~\cite{zheng2023temporal}, incorporate spatiotemporal graph structures for human mobility simulation~\cite{wang2024spatiotemporal}, and use Transformer-based transfer learning for cross-city trajectory generation~\cite{wang2024cola}. 
These methods highlight the importance of temporal evolution and transferable behavioral patterns, but mainly target mobility or general dynamic representation learning rather than label-scarce anomaly detection on rapidly evolving cryptocurrency transaction graphs.

\paragraph{Motif in Graph.}
Traditional GNNs aggregate pairwise connections and are limited by the expressive power of the 1-WL test~\cite{lee2019graph}. 
Graph motifs, as recurring higher-order subgraphs, have been used to enhance structural expressiveness, including motif-based attention~\cite{lee2019graph}, motif-augmented attributed networks~\cite{huang2021hybrid}, and motif-based GNNs~\cite{monti2018motifnet}. 
Nevertheless, most motif-based methods ignore temporal and directional information or require explicit graph augmentation, limiting their scalability on large dynamic transaction graphs.

\paragraph{Graph Test-time Adaptation.}
Test-time adaptation (TTA) adapts pre-trained models to distribution shifts during inference without labeled data.
Existing graph TTA methods rely on min-max optimization for test-domain adaptation~\cite{chen2022graphtta}, graph structure editing~\cite{jin2022empowering}, regularized prototype supervision~\cite{zhaotest}, edge-importance-based graph augmentation~\cite{zhang2024fully}, or homophily-based pseudo-label denoising~\cite{zhengtest}.
However, they mainly handle global distribution shifts and may overlook fine-grained transactional patterns crucial for cryptocurrency fraud detection.

\paragraph{Summary of Differences.}
Overall, existing studies provide valuable foundations but remain insufficient for blockchain fraud detection. 
GAD and graph foundation models mainly rely on static representations, motif-based GNNs often suffer from temporal or scalability limitations, and graph TTA methods usually overlook fine-grained higher-order transaction structures. 
Moreover, real-time intelligent big data processing emphasizes scalable analysis over continuously generated large-scale streams~\cite{zheng2019real}, which is important for practical blockchain monitoring. 
Different from these works, our approach jointly integrates motif-aware modeling with label-free test-time adaptation, enabling efficient adaptation to evolving higher-order transaction patterns in dynamic blockchain systems.

\section{Method}

\begin{figure*}[t]
    \centering
    \includegraphics[width=0.85\textwidth]{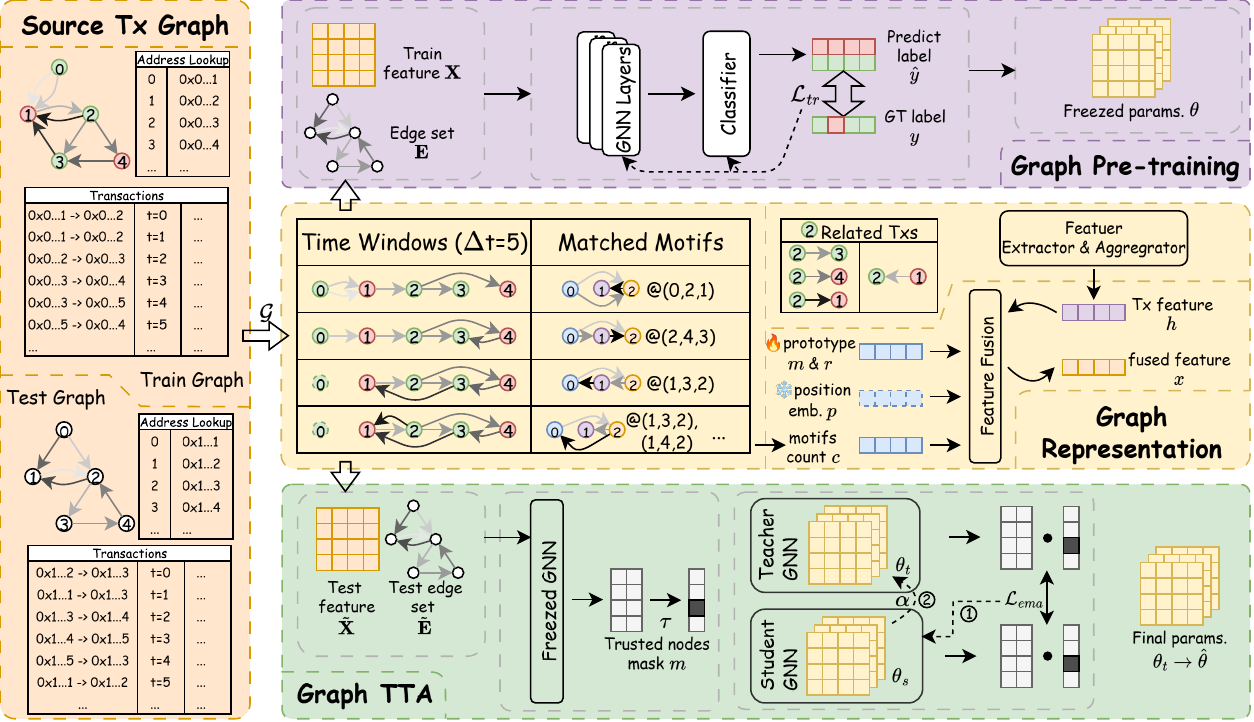}
    \caption{Overall framework of our proposed \NAME.}
    \label{fig:model}
\end{figure*}

\subsection{Blockchain Transaction Graph}

Given a cryptocurrency transaction set $\mathcal{R}=\{r_1,\ldots,r_M\}$, each transaction is denoted as $r_i=(s_i,e_i,t_i,c_i)$, where $s_i$, $e_i$, $t_i$, and $c_i$ denote the sender, receiver, timestamp, and amount, respectively.
We formulate $\mathcal{R}$ as a directed temporal multi-graph $\mathcal{G}=(V,E,\mathbf{X})$, where $V=\{s_i\}\cup\{e_i\}$ is the address set with $|V|=N$, $E=\{r_i\}_{i=1}^M$ is the transaction set, and $\mathbf{X}\in\mathbb{R}^{N\times d}$ denotes node features.
Our goal is to learn a GNN-based classifier $\hat{\mathbf{y}}_i=\mathbf{GNN}_{\theta}(v_i|\mathcal{G})$ to distinguish benign and anomalous addresses.

\subsection{Temporal Motif-aware Graph Representation}

\begin{algorithm}[t]
    \caption{Motif Matching}
    \label{alg:motifs}
    \textbf{Input}: Transaction set $\mathcal{R}$\\
    \textbf{Parameter}: Maximum time window $t_w$, edge limit $k$, aggregation range $\Delta t$\\
    \textbf{Output}: Motif count matrix $\mathbf{C}$
    \begin{algorithmic}[1]
        \STATE Sort transactions in $\mathcal{R}$ by timestamp
        \STATE Initialize motif counts $\mathbf{C}$
        \IF{time aggregation is enabled}
            \STATE Aggregate same-direction transactions within $[t_i-\Delta t,t_i]$ into $\tilde{\mathcal{R}}$
        \ELSE
            \STATE $\tilde{\mathcal{R}}=\mathcal{R}$
        \ENDIF
        \FOR{each transaction $r_i\in\mathcal{R}$}
            \STATE Sample at most $k$ historical transactions from $\tilde{\mathcal{R}}$ within $[t_i-t_w,t_i]$ as $\mathcal{S}_i$
            \FOR{each pair $(r_j,r_m)$ in $\mathcal{S}_i$ with $t_j<t_m<t_i$}
                \IF{$\{r_j,r_m,r_i\}$ forms a 3-node temporal motif}
                    \STATE Update the corresponding motif-role count in $\mathbf{C}$
                \ENDIF
            \ENDFOR
        \ENDFOR
        \STATE \textbf{return} $\mathbf{C}$
    \end{algorithmic}
\end{algorithm}

\begin{figure}[t]
    \centering
    \includegraphics[width=0.45\textwidth]{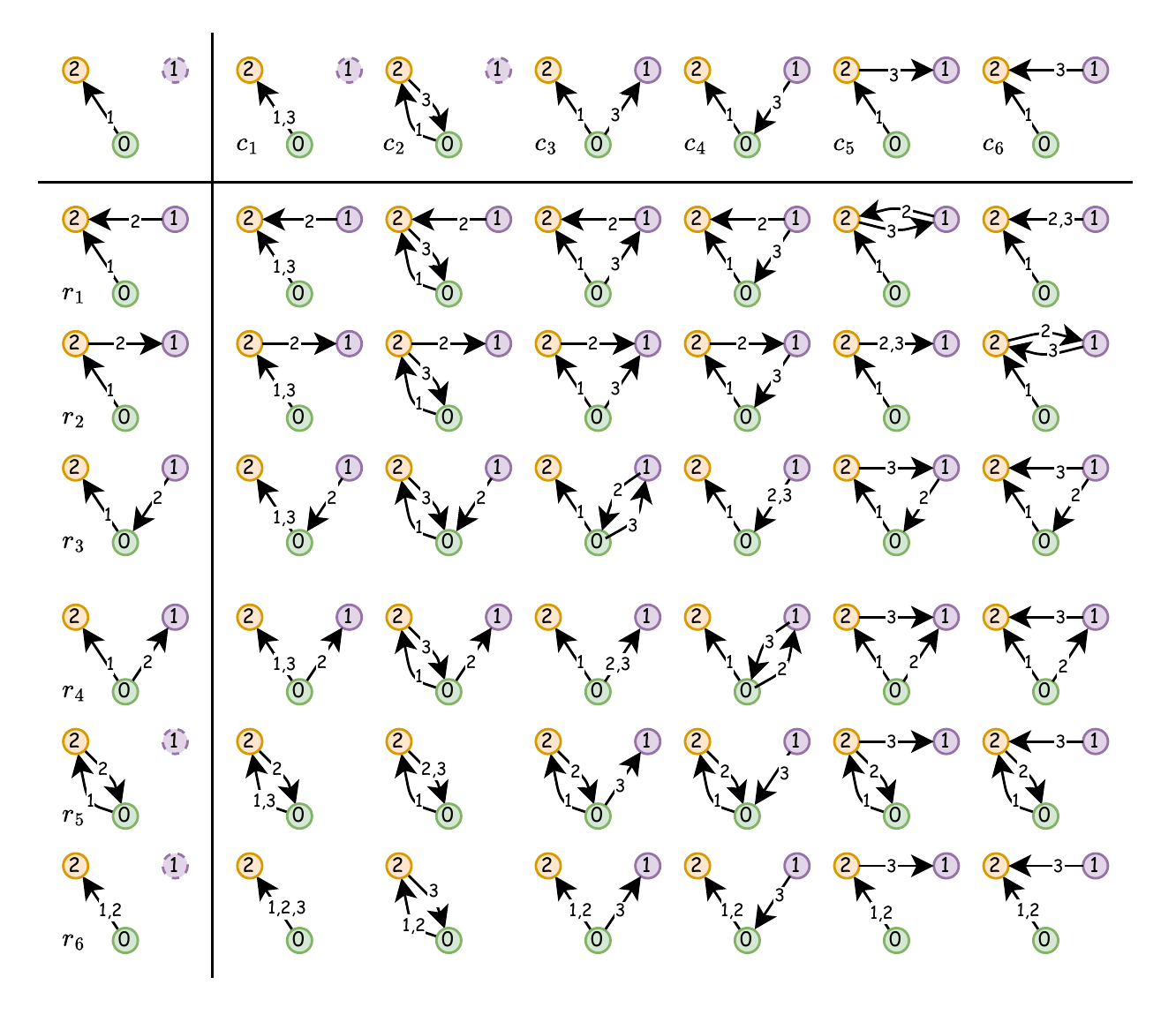}
    \caption{All possible directed motifs with 3 nodes and 3 edges.}
    \label{fig:motifs}
\end{figure}

\paragraph{Temporal Motif Computation.}
Existing GAD methods mainly rely on multi-hop graph aggregation and node-level classifiers, making it difficult to capture fine-grained intra- and inter-anomaly transaction patterns.
We therefore introduce temporal motifs to characterize higher-order transaction behaviors and enhance node representations.

Motifs are recurring subgraph patterns. 
In this work, we focus on 3-node and 3-edge temporal motifs, which capture interactions beyond pairwise connections while remaining computationally tractable~\cite{paranjape2017motifs}, as shown in Figure~\ref{fig:motifs}.
Two-node patterns can be captured by ordinary GNN aggregation, whereas four-node motifs are much more expensive and harder to represent compactly.
Algorithm~\ref{alg:motifs} shows our motif matching procedure.
With edge limitation and time aggregation, the complexity is reduced from $\mathcal{O}(M^3)$ to $\mathcal{O}(M\cdot k^2)$.
The resulting motif-role counts are stored in $\mathbf{C}\in\mathbb{R}^{N\times(3\times36)}$.

\paragraph{Temporal Motif Representation.}
Given 36 temporal motif types, we learn a shared motif prototype matrix $\mathbf{M}=[\mathbf{m}_1,\ldots,\mathbf{m}_{36}]$.
For each motif type $m_k$ and each node role $q$, we construct a role-aware motif embedding using the motif prototype, a learnable role embedding $\mathbf{r}_{k,q}$, and a positional encoding $\mathbf{p}_{k,q}$.
For a motif $m_k=(V',E')$, where $E'=\{e_j\}_{j=1}^3$ and $e_j=(e_j^{src},e_j^{dst},t_j)$, the positional encoding of node $v_q$ is defined as
\begin{equation}
    \mathbf{p}_{k,q} = \sum_{e_j\in E'} \mathds{1}(v_q\in e_j)\cdot
    \left(\mathcal{T}(t_j)+\mathcal{E}(v_q,e_j)\right),
\end{equation}
where $\mathcal{T}(t_j)=\sin(\omega t_j+\theta)$ encodes temporal order, and
$\mathcal{E}(v_q,e_j)=\mathbf{MLP}([\mathds{1}(v_q=e_j^{src}),\mathds{1}(v_q=e_j^{dst})])$ encodes edge-level source/destination roles.
The motif-role embedding is then obtained by
\begin{equation}
    \mathbf{h}_{k,q}^{m}=\mathbf{m}_k + \mathbf{r}_{k,q} + \mathbf{p}_{k,q}.
\end{equation}
We collect all motif-role embeddings into $\mathbf{H}_m\in\mathbb{R}^{(3\times36)\times d_h}$ and compute node-level motif features as
\begin{equation}
    \mathbf{X}_{motif}=\mathbf{C}\mathbf{H}_m.
\end{equation}

\paragraph{Temporal Motif-enhanced Node Representation.}
We fuse original node features with motif features by
\begin{equation}
    \mathbf{X}=\mathbf{MLP}\left(W_f\mathbf{X}_{orig}+\mathbf{X}_{motif}+b\right),
    \label{eq:fusion}
\end{equation}
where $W_f$ and $b$ are learnable parameters, and $\mathbf{X}$ denotes the fused node features.

\subsection{Graph Pre-training}

We divide learning into graph pre-training and test-time adaptation.
In pre-training, the motif-enhanced graph is fed into a GNN backbone, which can be GCN~\cite{kipf2016semi}, GraphSAGE~\cite{hamilton2017inductive}, SpaceGNN~\cite{dong2025spacegnn}, or DGAGNN~\cite{duan2024dga}.
The general GNN pipeline is written as
\begin{equation}
    \hat{\mathbf{y}}_i =
    \mathbf{MLP}\left(
    \mathbf{Update}^{L}
    \left(\{\mathbf{z}_{v_j}^{L-1}\mid v_j\in\{v_i\}\cup\mathcal{N}(v_i)\}\right)
    \right),
\end{equation}
where $\mathbf{Update}^{L}$ is the $L$-th layer update function, $\mathbf{z}_{v_i}^{L}$ is the representation of $v_i$, and $\mathbf{z}_{v_i}^{0}=\mathbf{x}_i\in\mathbf{X}$.
The model is optimized with binary cross-entropy:
\begin{equation}
    \mathcal{L}_{tr}=-\frac{1}{N}\sum_{i=1}^{N}
    \left[
    y_i\log(\hat{y}_i)+(1-y_i)\log(1-\hat{y}_i)
    \right].
\end{equation}
We optimize $\theta$ by back-propagation and apply early stopping based on validation AUC-PRC with tolerance $\tau_{tolerance}$.

\subsection{Graph Test-time Adaptation for Blockchain}

In test-time adaptation (TTA), we adapt the pre-trained model to an unlabeled test graph $\mathcal{G}_{test}$ whose distribution differs from the training graph.
We initialize two identical GNNs from the pre-trained model: a student $\mathbf{GNN}_{\theta}^{S}$ for adaptation and a teacher $\mathbf{GNN}_{\theta}^{T}$ for stable supervision.

To avoid over-confident pseudo supervision~\cite{wu2020conditional}, we select reliable but non-trivial nodes using a confidence mask:
\begin{equation}
    B_i=\mathds{1}(\tau_{low}<p_i\leq\tau_{high}),
\end{equation}
where $p_i=\max\operatorname{softmax}(\hat{\mathbf{y}}_i^T)$ is the teacher confidence, and $\mathcal{B}=\{i\mid B_i=1\}$.
At each adaptation step, we compute
\begin{equation}
    (\hat{\mathbf{Y}}^T,\mathbf{H}^T)=\mathbf{GNN}_{\theta}^{T}(\mathcal{G}_{test}),\quad
    (\hat{\mathbf{Y}}^S,\mathbf{H}^S)=\mathbf{GNN}_{\theta}^{S}(\tilde{\mathcal{G}}_{test}),
\end{equation}
where $\tilde{\mathcal{G}}_{test}$ is obtained by random edge dropping to produce a semantic-preserving perturbation.
We regularize the student by matching teacher and student embeddings:
\begin{equation}
    \mathcal{L}_{sim}
    =\frac{1}{|\mathcal{B}|}\sum_{i\in\mathcal{B}}
    \left(1-\cos(\mathbf{h}_i^T,\mathbf{h}_i^S)\right).
\end{equation}

To improve robustness to unstable blockchain patterns, we further adopt an InfoNCE loss~\cite{oord2018representation}:
\begin{equation}
    \mathcal{L}_{Info}
    =
    -\frac{1}{|\mathcal{B}|}
    \sum_{i\in\mathcal{B}}
    \log
    \frac{\operatorname{sim}(\mathbf{h}_i,\mathbf{h}_i^+)}
    {\operatorname{sim}(\mathbf{h}_i,\mathbf{h}_i^+)
    +\sum_{\mathbf{h}_k^-\in\mathcal{N}_i^-}
    \operatorname{sim}(\mathbf{h}_i,\mathbf{h}_k^-)},
\end{equation}
where $\operatorname{sim}(\mathbf{h},\mathbf{h}')=\exp(\cos(\mathbf{h},\mathbf{h}')/\tau)$.
For each node $v_i$, positives are sampled from nodes with similar motif features, i.e., $\cos(\mathbf{x}_{m_i},\mathbf{x}_{m_j})>\gamma$, while the remaining sampled nodes are treated as negatives $\mathcal{N}_i^-$.
The final TTA objective is
\begin{equation}
    \mathcal{L}_{TTA}=\mathcal{L}_{sim}+\beta\mathcal{L}_{Info},
\end{equation}
where $\beta$ controls the contrastive regularization strength.
We update the student parameters $\theta^S$ by back-propagation and update the teacher by exponential moving average:
\begin{equation}
    \theta^T=\alpha\theta^T+(1-\alpha)\theta^S.
\end{equation}
With $\alpha\in[0.9,0.999]$, the teacher preserves source-domain knowledge while gradually tracking the adapted student, reducing catastrophic forgetting during TTA.
During each training progress, $\alpha$ is fixed.

\section{Experiments}

\subsection{Experiment Setting}

\begin{table}[t]
    \centering
    \scalebox{0.84}{
        \begin{tabular}{lrrrr}
    \toprule
    Dataset & \#Nodes & \#Edges & \%Anomaly & Description \\
    \midrule
        Alpha & 115,488 & 897,308 & 8.07 & Exploitation \\
        Crypto & 222,761 & 835,302 & 4.38 & Hacker Attack \\
        Plus & 38,327 & 93,493 & 80.34 & Ponzi Scheme \\
        Upbit & 577,994 & 1,213,049 & 3.24 & Hacker Attack \\
        Trace & 1,698,331 & 2,969,691 & 0.16 & Money Laundering \\
    \bottomrule
\end{tabular}
    }
    \caption{Statistics of blockchain anomaly detection datasets.}
    \label{tab:dataset}
\end{table}

\paragraph{Datasets.}
We conduct experiments on five real-world blockchain anomaly detection datasets, as shown in Table~\ref{tab:dataset}. 
AlphaHomora (Alpha), CryptopiaHacker (Crypto), PlusTokenPonzi (Plus), and UpbitHack (Upbit) are from~\cite{lin2024denseflow}, while our private dataset Trace is collected through collaboration with our real-world partner.

\paragraph{Baselines.}
We evaluate \NAME~ with several GNN backbones, including Graph Convolutional Network (GCN)~\cite{kipf2016semi}, GraphSAGE (SAGE)~\cite{hamilton2017inductive}, SpaceGNN (SGNN)~\cite{dong2025spacegnn}, and DGAGNN (DGA)~\cite{duan2024dga}. 
We use GADBench~\cite{tang2023gadbench} implementations for the first three models and the official implementation for DGAGNN.
We also attempted temporal GNN baselines such as DyGFormer~\cite{yu2023towards} and SALoM~\cite{liu2026salom}, but their training cost is prohibitive on blockchain transaction graphs, with several runs exceeding 24 hours on public datasets and becoming harder on Trace. 
Following prior GAD studies, we therefore use representative GNN-based anomaly detection backbones for the main comparison and focus on the consistent gains brought by \NAME.

\paragraph{Evaluation protocol.}
To prevent data leakage, we record the activation time of each anomalous node and split the training, validation, and test sets chronologically with a ratio of 6:2:2.
Due to severe class imbalance, we do not use accuracy. 
Instead, we evaluate anomaly detection performance using Area Under the Precision-Recall Curve (AUC-PRC), Recall@$k$ (Rec@$k$), and F1-score. 
AUC-PRC measures the overall ranking quality under imbalance, Rec@$k$ evaluates the coverage of high-risk nodes in practical top-$k$ inspection, and F1-score reflects the balance between precision and recall under a decision threshold of 0.5. 
We set $k$ as the number of anomalous nodes in the test set.

\paragraph{Implementation details.}
We train each model for 200 epochs with early stopping based on validation AUC-PRC and run 10 epochs for TTA.
Unless otherwise specified, motif extraction uses $k=100$ and $\Delta t=3600$.
Since TTA requires hidden embeddings on the test graph, we modify all backbones to output the hidden embeddings before the classifier.
For DGAGNN, we disable its source graph oriented super mask during TTA.
All experiments are conducted on 2 Intel Xeon Platinum 8260L CPUs, 256 GiB RAM, and an NVIDIA GeForce RTX 4090 GPU with 24 GiB VRAM.

\subsection{Overall Comparison}

\begin{table*}[!htbp]
    \centering
    \scalebox{0.55}{
        \begin{tabular}{lll|rrrr|rrrr|rrrr}
    \toprule
    \multirow{2}{*}{GNN} & \multirow{2}{*}{Metric} & \multirow{2}*{\begin{tabular}{cc}
        \multirow{2}{*}{TTA} & Tr. \\
         & Te.
    \end{tabular}} & \multicolumn{4}{|c}{Alpha} & \multicolumn{4}{|c}{Crypto} & \multicolumn{4}{|c}{Plus} \\
    \cmidrule{4-15}
    & & & Crypto & Plus & Upbit & Trace & Alpha & Plus & Upbit & Trace & Alpha & Crypto & Upbit & Trace \\
    \midrule
    \multirow{9}{*}{GCN} & \multirow{2}{*}{AUC-PRC} & Origin & 0.2292 & 0.6241 & 0.1752 & 0.0047 & 0.5552 & 0.6333 & 0.1101 & 0.0022 & 0.1490 & 0.0562 & 0.0974 & 0.0012 \\
    & & Ours & \cellcolor{gray!15} \makecell[r]{\textbf{0.3636}\\{\tiny $\pm$0.0008 $**$}}  & \cellcolor{gray!15} \makecell[r]{\textbf{0.7647}\\{\tiny $\pm$0.0128 $*$}}  & \cellcolor{gray!15} \makecell[r]{\textbf{0.4457}\\{\tiny $\pm$0.0016 $**$}}  & \cellcolor{gray!15} \makecell[r]{\textbf{0.0399}\\{\tiny $\pm$0.0002 $**$}}  & \cellcolor{gray!15} \makecell[r]{\textbf{0.6439}\\{\tiny $\pm$0.0006 $**$}}  & \cellcolor{gray!15} \makecell[r]{\textbf{0.8159}\\{\tiny $\pm$0.0169 $*$}}  & \cellcolor{gray!15} \makecell[r]{\textbf{0.3037}\\{\tiny $\pm$0.0007 $**$}}  & \cellcolor{gray!15} \makecell[r]{\textbf{0.0060}\\{\tiny $\pm$0.0000 $**$}}  & \cellcolor{gray!15} \makecell[r]{\textbf{0.1673}\\{\tiny $\pm$0.0005 $**$}}  & \cellcolor{gray!15} \makecell[r]{\textbf{0.0613}\\{\tiny $\pm$0.0001 $**$}}  & \cellcolor{gray!15} \makecell[r]{\textbf{0.1013}\\{\tiny $\pm$0.0001 $**$}}  & \cellcolor{gray!15} \makecell[r]{\textbf{0.0012}\\{\tiny $\pm$0.0000 $**$}} \\
    \cmidrule{2-15}
     & \multirow{2}{*}{Rec@$k$} & Origin & 0.2838 & 0.7562 & 0.1961 & 0.0000 & 0.4978 & 0.7576 & 0.2323 & \textbf{0.0000} & \textbf{0.3246} & \textbf{0.2274} & \textbf{0.3049} & 0.0007 \\
    & & Ours & \cellcolor{gray!15} \makecell[r]{\textbf{0.3466}\\{\tiny $\pm$0.0011 $**$}}  & \cellcolor{gray!15} \makecell[r]{\textbf{0.7680}\\{\tiny $\pm$0.0000 $**$}}  & \cellcolor{gray!15} \makecell[r]{\textbf{0.4705}\\{\tiny $\pm$0.0005 $**$}}  & \cellcolor{gray!15} \makecell[r]{\textbf{0.1044}\\{\tiny $\pm$0.0002 $**$}}  & \cellcolor{gray!15} \makecell[r]{\textbf{0.5833}\\{\tiny $\pm$0.0001 $**$}}  & \cellcolor{gray!15} \makecell[r]{\textbf{0.7624}\\{\tiny $\pm$0.0001 $**$}}  & \cellcolor{gray!15} \makecell[r]{\textbf{0.4319}\\{\tiny $\pm$0.0002 $**$}}  & \cellcolor{gray!15} \makecell[r]{\textbf{0.0000}\\{\tiny $\pm$0.0000}}  & \cellcolor{gray!15} \makecell[r]{0.0000\\{\tiny $\pm$0.0001}}  & \cellcolor{gray!15} \makecell[r]{0.0000\\{\tiny $\pm$0.0000}}  & \cellcolor{gray!15} \makecell[r]{0.0000\\{\tiny $\pm$0.0000}}  & \cellcolor{gray!15} \makecell[r]{\textbf{0.0018}\\{\tiny $\pm$0.0000 $**$}} \\
    \cmidrule{2-15}
     & \multirow{2}{*}{F1} & Origin & 0.3062 & \textbf{0.8910} & 0.4183 & 0.0106 & 0.5354 & 0.8910 & 0.2929 & 0.0044 & 0.2516 & 0.1154 & 0.1844 & 0.0032 \\
    & & Ours & \cellcolor{gray!15} \makecell[r]{\textbf{0.3787}\\{\tiny $\pm$0.0010 $**$}}  & \cellcolor{gray!15} \makecell[r]{0.8910\\{\tiny $\pm$0.0000}}  & \cellcolor{gray!15} \makecell[r]{\textbf{0.5004}\\{\tiny $\pm$0.0004 $**$}}  & \cellcolor{gray!15} \makecell[r]{\textbf{0.1461}\\{\tiny $\pm$0.0004 $**$}}  & \cellcolor{gray!15} \makecell[r]{\textbf{0.5964}\\{\tiny $\pm$0.0004 $**$}}  & \cellcolor{gray!15} \makecell[r]{\textbf{0.8911}\\{\tiny $\pm$0.0000 $**$}}  & \cellcolor{gray!15} \makecell[r]{\textbf{0.4422}\\{\tiny $\pm$0.0009 $**$}}  & \cellcolor{gray!15} \makecell[r]{\textbf{0.0567}\\{\tiny $\pm$0.0002 $**$}}  & \cellcolor{gray!15} \makecell[r]{\textbf{0.2855}\\{\tiny $\pm$0.0008 $**$}}  & \cellcolor{gray!15} \makecell[r]{\textbf{0.1260}\\{\tiny $\pm$0.0000 $**$}}  & \cellcolor{gray!15} \makecell[r]{\textbf{0.1922}\\{\tiny $\pm$0.0001 $**$}}  & \cellcolor{gray!15} \makecell[r]{\textbf{0.0034}\\{\tiny $\pm$0.0000 $**$}} \\
    \midrule
    \multirow{9}{*}{SAGE} & \multirow{2}{*}{AUC-PRC} & Origin & 0.1946 & 0.8278 & 0.0664 & 0.0021 & 0.3541 & 0.6145 & \textbf{0.1047} & \textbf{0.0024} & 0.1093 & 0.0456 & 0.0503 & \textbf{0.0017} \\
    & & Ours & \cellcolor{gray!15} \makecell[r]{\textbf{0.2080}\\{\tiny $\pm$0.0007 $*$}}  & \cellcolor{gray!15} \makecell[r]{\textbf{0.9985}\\{\tiny $\pm$0.0000 $**$}}  & \cellcolor{gray!15} \makecell[r]{\textbf{0.2429}\\{\tiny $\pm$0.0011 $**$}}  & \cellcolor{gray!15} \makecell[r]{\textbf{0.0023}\\{\tiny $\pm$0.0000 $**$}}  & \cellcolor{gray!15} \makecell[r]{\textbf{0.4880}\\{\tiny $\pm$0.0016 $**$}}  & \cellcolor{gray!15} \makecell[r]{\textbf{0.9907}\\{\tiny $\pm$0.0000 $**$}}  & \cellcolor{gray!15} \makecell[r]{0.0832\\{\tiny $\pm$0.0007}}  & \cellcolor{gray!15} \makecell[r]{0.0015\\{\tiny $\pm$0.0000}}  & \cellcolor{gray!15} \makecell[r]{\textbf{0.1195}\\{\tiny $\pm$0.0000 $**$}}  & \cellcolor{gray!15} \makecell[r]{\textbf{0.0557}\\{\tiny $\pm$0.0000 $**$}}  & \cellcolor{gray!15} \makecell[r]{\textbf{0.0777}\\{\tiny $\pm$0.0000 $**$}}  & \cellcolor{gray!15} \makecell[r]{0.0017\\{\tiny $\pm$0.0000}} \\
    \cmidrule{2-15}
     & \multirow{2}{*}{Rec@$k$} & Origin & 0.2269 & 0.9396 & 0.0000 & 0.0018 & 0.4567 & 0.7690 & \textbf{0.0616} & \textbf{0.0089} & 0.0000 & \textbf{0.0000} & \textbf{0.0000} & \textbf{0.0000} \\
    & & Ours & \cellcolor{gray!15} \makecell[r]{\textbf{0.2847}\\{\tiny $\pm$0.0057 $*$}}  & \cellcolor{gray!15} \makecell[r]{\textbf{0.9926}\\{\tiny $\pm$0.0002 $**$}}  & \cellcolor{gray!15} \makecell[r]{\textbf{0.3843}\\{\tiny $\pm$0.0007 $**$}}  & \cellcolor{gray!15} \makecell[r]{\textbf{0.0111}\\{\tiny $\pm$0.0000 $**$}}  & \cellcolor{gray!15} \makecell[r]{\textbf{0.5400}\\{\tiny $\pm$0.0029 $**$}}  & \cellcolor{gray!15} \makecell[r]{\textbf{0.9855}\\{\tiny $\pm$0.0000 $**$}}  & \cellcolor{gray!15} \makecell[r]{0.0145\\{\tiny $\pm$0.0034}}  & \cellcolor{gray!15} \makecell[r]{0.0000\\{\tiny $\pm$0.0000}}  & \cellcolor{gray!15} \makecell[r]{\textbf{0.0001}\\{\tiny $\pm$0.0000 $**$}}  & \cellcolor{gray!15} \makecell[r]{\textbf{0.0000}\\{\tiny $\pm$0.0000}}  & \cellcolor{gray!15} \makecell[r]{\textbf{0.0000}\\{\tiny $\pm$0.0000}}  & \cellcolor{gray!15} \makecell[r]{\textbf{0.0000}\\{\tiny $\pm$0.0000}} \\
    \cmidrule{2-15}
     & \multirow{2}{*}{F1} & Origin & 0.4004 & 0.9924 & 0.3588 & 0.0050 & 0.4734 & 0.8970 & 0.1654 & 0.0032 & 0.2119 & \textbf{0.1119} & 0.1354 & 0.0036 \\
    & & Ours & \cellcolor{gray!15} \makecell[r]{\textbf{0.4138}\\{\tiny $\pm$0.0033 $*$}}  & \cellcolor{gray!15} \makecell[r]{\textbf{0.9940}\\{\tiny $\pm$0.0002 $*$}}  & \cellcolor{gray!15} \makecell[r]{\textbf{0.3975}\\{\tiny $\pm$0.0009 $**$}}  & \cellcolor{gray!15} \makecell[r]{\textbf{0.0147}\\{\tiny $\pm$0.0000 $**$}}  & \cellcolor{gray!15} \makecell[r]{\textbf{0.5518}\\{\tiny $\pm$0.0017 $**$}}  & \cellcolor{gray!15} \makecell[r]{\textbf{0.9925}\\{\tiny $\pm$0.0000 $**$}}  & \cellcolor{gray!15} \makecell[r]{\textbf{0.1808}\\{\tiny $\pm$0.0006 $**$}}  & \cellcolor{gray!15} \makecell[r]{\textbf{0.0037}\\{\tiny $\pm$0.0000 $**$}}  & \cellcolor{gray!15} \makecell[r]{\textbf{0.2139}\\{\tiny $\pm$0.0000 $**$}}  & \cellcolor{gray!15} \makecell[r]{0.1117\\{\tiny $\pm$0.0000}}  & \cellcolor{gray!15} \makecell[r]{\textbf{0.1460}\\{\tiny $\pm$0.0000 $**$}}  & \cellcolor{gray!15} \makecell[r]{\textbf{0.0039}\\{\tiny $\pm$0.0000 $**$}} \\
    \midrule
    \multirow{9}{*}{DGA} & \multirow{2}{*}{AUC-PRC} & Origin & 0.2181 & 0.7195 & 0.1121 & 0.0020 & 0.3709 & 0.9869 & 0.1331 & 0.0017 & 0.0652 & 0.0705 & 0.0343 & 0.0053 \\
    & & Ours & \cellcolor{gray!15} \makecell[r]{\textbf{0.4152}\\{\tiny $\pm$0.0010 $**$}}  & \cellcolor{gray!15} \makecell[r]{\textbf{0.9931}\\{\tiny $\pm$0.0002 $**$}}  & \cellcolor{gray!15} \makecell[r]{\textbf{0.4510}\\{\tiny $\pm$0.0046 $**$}}  & \cellcolor{gray!15} \makecell[r]{\textbf{0.0174}\\{\tiny $\pm$0.0001 $**$}}  & \cellcolor{gray!15} \makecell[r]{\textbf{0.5100}\\{\tiny $\pm$0.0079 $*$}}  & \cellcolor{gray!15} \makecell[r]{\textbf{0.9922}\\{\tiny $\pm$0.0000 $**$}}  & \cellcolor{gray!15} \makecell[r]{\textbf{0.2153}\\{\tiny $\pm$0.0004 $**$}}  & \cellcolor{gray!15} \makecell[r]{\textbf{0.0184}\\{\tiny $\pm$0.0009 $*$}}  & \cellcolor{gray!15} \makecell[r]{\textbf{0.0984}\\{\tiny $\pm$0.0035 $*$}}  & \cellcolor{gray!15} \makecell[r]{\textbf{0.1017}\\{\tiny $\pm$0.0021 $*$}}  & \cellcolor{gray!15} \makecell[r]{\textbf{0.2802}\\{\tiny $\pm$0.0137 $*$}}  & \cellcolor{gray!15} \makecell[r]{\textbf{0.0091}\\{\tiny $\pm$0.0001 $**$}} \\
    \cmidrule{2-15}
     & \multirow{2}{*}{Rec@$k$} & Origin & 0.2883 & 0.7564 & 0.2334 & 0.0037 & 0.3840 & 0.9383 & 0.2890 & 0.0000 & 0.0677 & 0.0845 & 0.0384 & 0.0126 \\
    & & Ours & \cellcolor{gray!15} \makecell[r]{\textbf{0.4125}\\{\tiny $\pm$0.0029 $**$}}  & \cellcolor{gray!15} \makecell[r]{\textbf{0.9819}\\{\tiny $\pm$0.0017 $**$}}  & \cellcolor{gray!15} \makecell[r]{\textbf{0.4313}\\{\tiny $\pm$0.0022 $**$}}  & \cellcolor{gray!15} \makecell[r]{\textbf{0.0334}\\{\tiny $\pm$0.0002 $**$}}  & \cellcolor{gray!15} \makecell[r]{\textbf{0.5263}\\{\tiny $\pm$0.0071 $*$}}  & \cellcolor{gray!15} \makecell[r]{\textbf{0.9852}\\{\tiny $\pm$0.0000 $**$}}  & \cellcolor{gray!15} \makecell[r]{\textbf{0.3872}\\{\tiny $\pm$0.0051 $*$}}  & \cellcolor{gray!15} \makecell[r]{\textbf{0.0435}\\{\tiny $\pm$0.0011 $**$}}  & \cellcolor{gray!15} \makecell[r]{\textbf{0.0813}\\{\tiny $\pm$0.0097}}  & \cellcolor{gray!15} \makecell[r]{\textbf{0.1635}\\{\tiny $\pm$0.0017 $**$}}  & \cellcolor{gray!15} \makecell[r]{\textbf{0.2792}\\{\tiny $\pm$0.0104 $**$}}  & \cellcolor{gray!15} \makecell[r]{\textbf{0.0479}\\{\tiny $\pm$0.0003 $**$}} \\
    \cmidrule{2-15}
     & \multirow{2}{*}{F1} & Origin & 0.3399 & 0.8910 & 0.3962 & 0.0212 & 0.3592 & 0.9897 & 0.2826 & 0.0032 & 0.1530 & 0.0945 & 0.0632 & 0.0056 \\
    & & Ours & \cellcolor{gray!15} \makecell[r]{\textbf{0.4413}\\{\tiny $\pm$0.0005 $**$}}  & \cellcolor{gray!15} \makecell[r]{\textbf{0.9825}\\{\tiny $\pm$0.0013 $**$}}  & \cellcolor{gray!15} \makecell[r]{\textbf{0.4684}\\{\tiny $\pm$0.0023 $**$}}  & \cellcolor{gray!15} \makecell[r]{\textbf{0.0568}\\{\tiny $\pm$0.0007 $**$}}  & \cellcolor{gray!15} \makecell[r]{\textbf{0.5334}\\{\tiny $\pm$0.0072 $**$}}  & \cellcolor{gray!15} \makecell[r]{\textbf{0.9925}\\{\tiny $\pm$0.0000 $**$}}  & \cellcolor{gray!15} \makecell[r]{\textbf{0.4141}\\{\tiny $\pm$0.0053 $**$}}  & \cellcolor{gray!15} \makecell[r]{\textbf{0.0700}\\{\tiny $\pm$0.0029 $**$}}  & \cellcolor{gray!15} \makecell[r]{\textbf{0.2056}\\{\tiny $\pm$0.0026 $*$}}  & \cellcolor{gray!15} \makecell[r]{\textbf{0.1888}\\{\tiny $\pm$0.0009 $**$}}  & \cellcolor{gray!15} \makecell[r]{\textbf{0.3244}\\{\tiny $\pm$0.0165 $*$}}  & \cellcolor{gray!15} \makecell[r]{\textbf{0.0558}\\{\tiny $\pm$0.0002 $**$}} \\
    \midrule
    \multirow{9}{*}{SGNN} & \multirow{2}{*}{AUC-PRC} & Origin & 0.5077 & \textbf{0.9963} & 0.3772 & \textbf{0.0256} & \textbf{0.6418} & 0.9895 & \textbf{0.7074} & \textbf{0.0033} & 0.2108 & 0.1065 & 0.2055 & \textbf{0.0054} \\
    & & Ours & \cellcolor{gray!15} \makecell[r]{\textbf{0.5904}\\{\tiny $\pm$0.0001 $**$}}  & \cellcolor{gray!15} \makecell[r]{0.9956\\{\tiny $\pm$0.0000}}  & \cellcolor{gray!15} \makecell[r]{\textbf{0.7443}\\{\tiny $\pm$0.0003 $**$}}  & \cellcolor{gray!15} -  & \cellcolor{gray!15} \makecell[r]{0.6094\\{\tiny $\pm$0.0005}}  & \cellcolor{gray!15} \makecell[r]{\textbf{0.9974}\\{\tiny $\pm$0.0000 $**$}}  & \cellcolor{gray!15} \makecell[r]{0.6972\\{\tiny $\pm$0.0007}}  & \cellcolor{gray!15} -  & \cellcolor{gray!15} \makecell[r]{\textbf{0.2221}\\{\tiny $\pm$0.0000 $**$}}  & \cellcolor{gray!15} \makecell[r]{\textbf{0.1324}\\{\tiny $\pm$0.0000 $**$}}  & \cellcolor{gray!15} \makecell[r]{\textbf{0.3398}\\{\tiny $\pm$0.0002 $**$}}  & \cellcolor{gray!15} - \\
    \cmidrule{2-15}
     & \multirow{2}{*}{Rec@$k$} & Origin & 0.4915 & 0.9915 & 0.4070 & \textbf{0.0588} & \textbf{0.6374} & 0.9795 & \textbf{0.6693} & \textbf{0.0052} & \textbf{0.0000} & \textbf{0.0000} & 0.0000 & \textbf{0.0044} \\
    & & Ours & \cellcolor{gray!15} \makecell[r]{\textbf{0.5486}\\{\tiny $\pm$0.0007 $**$}}  & \cellcolor{gray!15} \makecell[r]{\textbf{0.9937}\\{\tiny $\pm$0.0001 $**$}}  & \cellcolor{gray!15} \makecell[r]{\textbf{0.6698}\\{\tiny $\pm$0.0000 $**$}}  & \cellcolor{gray!15} -  & \cellcolor{gray!15} \makecell[r]{0.5832\\{\tiny $\pm$0.0006}}  & \cellcolor{gray!15} \makecell[r]{\textbf{0.9875}\\{\tiny $\pm$0.0000 $**$}}  & \cellcolor{gray!15} \makecell[r]{0.6470\\{\tiny $\pm$0.0007}}  & \cellcolor{gray!15} -  & \cellcolor{gray!15} \makecell[r]{\textbf{0.0000}\\{\tiny $\pm$0.0000}}  & \cellcolor{gray!15} \makecell[r]{\textbf{0.0000}\\{\tiny $\pm$0.0000}}  & \cellcolor{gray!15} \makecell[r]{\textbf{0.2857}\\{\tiny $\pm$0.0554 $*$}}  & \cellcolor{gray!15} - \\
    \cmidrule{2-15}
     & \multirow{2}{*}{F1} & Origin & 0.5747 & 0.9917 & 0.6165 & \textbf{0.0162} & 0.5535 & 0.9440 & 0.5199 & \textbf{0.0087} & \textbf{0.3842} & \textbf{0.2529} & 0.4727 & \textbf{0.0899} \\
    & & Ours & \cellcolor{gray!15} \makecell[r]{\textbf{0.5763}\\{\tiny $\pm$0.0004 $*$}}  & \cellcolor{gray!15} \makecell[r]{\textbf{0.9942}\\{\tiny $\pm$0.0001 $**$}}  & \cellcolor{gray!15} \makecell[r]{\textbf{0.6933}\\{\tiny $\pm$0.0002 $**$}}  & \cellcolor{gray!15} -  & \cellcolor{gray!15} \makecell[r]{\textbf{0.5852}\\{\tiny $\pm$0.0004 $**$}}  & \cellcolor{gray!15} \makecell[r]{\textbf{0.9921}\\{\tiny $\pm$0.0000 $**$}}  & \cellcolor{gray!15} \makecell[r]{\textbf{0.6561}\\{\tiny $\pm$0.0005 $**$}}  & \cellcolor{gray!15} -  & \cellcolor{gray!15} \makecell[r]{0.3674\\{\tiny $\pm$0.0000}}  & \cellcolor{gray!15} \makecell[r]{0.2397\\{\tiny $\pm$0.0001}}  & \cellcolor{gray!15} \makecell[r]{\textbf{0.5009}\\{\tiny $\pm$0.0001 $**$}}  & \cellcolor{gray!15} - \\
    \bottomrule
\end{tabular}
    }
    \caption{Overall comparison results. Models are pre-trained on AlphaHomora, CryptopiaHacker, and PlusTokenPonzi. $**$ indicates $p<0.001$ and $*$ indicates $p<0.1$ in the paired t-test compared with the baseline.}
    \label{tab:tta_comp_1}
\end{table*}

\begin{table*}[!htbp]
    \centering
    \scalebox{0.6}{
        \begin{tabular}{lll|rrrr|rrrr}
    \toprule
    \multirow{2}{*}{GNN} & \multirow{2}{*}{Metric} & \multirow{2}*{\begin{tabular}{cc}
        \multirow{2}{*}{TTA} & Tr. \\
         & Te.
    \end{tabular}} & \multicolumn{4}{|c}{Upbit} & \multicolumn{4}{|c}{Trace} \\
    \cmidrule{4-11}
    & & & Alpha & Crypto & Plus & Trace & Alpha & Crypto & Plus & Upbit \\
    \midrule
    \multirow{6}{*}{GCN} & \multirow{2}{*}{AUC-PRC} & Origin & 0.4164 & 0.2628 & 0.8734 & 0.0378 & 0.4211 & 0.3006 & 0.9608 & 0.1222 \\
    & & Ours & \cellcolor{gray!15} \makecell[r]{\textbf{0.5471}\\{\tiny $\pm$0.0002 $**$}} & \cellcolor{gray!15} \makecell[r]{\textbf{0.4274}\\{\tiny $\pm$0.0004 $**$}} & \cellcolor{gray!15} \makecell[r]{\textbf{0.9874}\\{\tiny $\pm$0.0000 $**$}} & \cellcolor{gray!15} \makecell[r]{\textbf{0.0386}\\{\tiny $\pm$0.0000 $**$}} & \cellcolor{gray!15} \makecell[r]{\textbf{0.5177}\\{\tiny $\pm$0.0002 $**$}} & \cellcolor{gray!15} \makecell[r]{\textbf{0.3265}\\{\tiny $\pm$0.0001 $**$}} & \cellcolor{gray!15} \makecell[r]{\textbf{0.9619}\\{\tiny $\pm$0.0004 $*$}} & \cellcolor{gray!15} \makecell[r]{\textbf{0.1442}\\{\tiny $\pm$0.0002 $**$}} \\
    \cmidrule{2-11}
     & \multirow{2}{*}{Rec@$k$} & Origin & 0.3606 & 0.2285 & 0.7562 & 0.1035 & 0.4018 & 0.3334 & 0.8834 & 0.1142 \\
    & & Ours & \cellcolor{gray!15} \makecell[r]{\textbf{0.5088}\\{\tiny $\pm$0.0007 $**$}} & \cellcolor{gray!15} \makecell[r]{\textbf{0.4156}\\{\tiny $\pm$0.0002 $**$}} & \cellcolor{gray!15} \makecell[r]{\textbf{0.9768}\\{\tiny $\pm$0.0002 $**$}} & \cellcolor{gray!15} \makecell[r]{\textbf{0.1046}\\{\tiny $\pm$0.0000 $**$}} & \cellcolor{gray!15} \makecell[r]{\textbf{0.4899}\\{\tiny $\pm$0.0002 $**$}} & \cellcolor{gray!15} \makecell[r]{\textbf{0.3486}\\{\tiny $\pm$0.0003 $**$}} & \cellcolor{gray!15} \makecell[r]{\textbf{0.8848}\\{\tiny $\pm$0.0017}} & \cellcolor{gray!15} \makecell[r]{\textbf{0.1478}\\{\tiny $\pm$0.0001 $**$}} \\
    \cmidrule{2-11}
     & \multirow{2}{*}{F1} & Origin & 0.4542 & 0.3323 & 0.9563 & 0.0611 & 0.4666 & 0.3722 & 0.8910 & 0.1742 \\
    & & Ours & \cellcolor{gray!15} \makecell[r]{\textbf{0.5220}\\{\tiny $\pm$0.0005 $**$}} & \cellcolor{gray!15} \makecell[r]{\textbf{0.4274}\\{\tiny $\pm$0.0007 $**$}} & \cellcolor{gray!15} \makecell[r]{\textbf{0.9846}\\{\tiny $\pm$0.0000 $**$}} & \cellcolor{gray!15} \makecell[r]{\textbf{0.1572}\\{\tiny $\pm$0.0000 $**$}} & \cellcolor{gray!15} \makecell[r]{\textbf{0.5524}\\{\tiny $\pm$0.0003 $**$}} & \cellcolor{gray!15} \makecell[r]{\textbf{0.3991}\\{\tiny $\pm$0.0003 $**$}} & \cellcolor{gray!15} \makecell[r]{\textbf{0.9049}\\{\tiny $\pm$0.0010 $*$}} & \cellcolor{gray!15} \makecell[r]{\textbf{0.2096}\\{\tiny $\pm$0.0003 $**$}} \\
    \midrule
    \multirow{6}{*}{SAGE} & \multirow{2}{*}{AUC-PRC} & Origin & 0.1689 & 0.1346 & 0.9214 & 0.0012 & \textbf{0.0533} & \textbf{0.0359} & \textbf{0.9888} & 0.0190 \\
    & & Ours & \cellcolor{gray!15} \makecell[r]{\textbf{0.3908}\\{\tiny $\pm$0.0014 $**$}} & \cellcolor{gray!15} \makecell[r]{\textbf{0.1918}\\{\tiny $\pm$0.0006 $**$}} & \cellcolor{gray!15} \makecell[r]{\textbf{0.9961}\\{\tiny $\pm$0.0000 $**$}} & \cellcolor{gray!15} \makecell[r]{\textbf{0.0042}\\{\tiny $\pm$0.0000 $**$}} & \cellcolor{gray!15} \makecell[r]{0.0532\\{\tiny $\pm$0.0000}} & \cellcolor{gray!15} \makecell[r]{0.0359\\{\tiny $\pm$0.0000}} & \cellcolor{gray!15} \makecell[r]{0.9876\\{\tiny $\pm$0.0000}} & \cellcolor{gray!15} \makecell[r]{\textbf{0.0191}\\{\tiny $\pm$0.0001 $*$}} \\
    \cmidrule{2-11}
     & \multirow{2}{*}{Rec@$k$} & Origin & 0.2358 & 0.2512 & 0.9380 & 0.0000 & 0.0209 & \textbf{0.0207} & 0.9853 & \textbf{0.0279} \\
    & & Ours & \cellcolor{gray!15} \makecell[r]{\textbf{0.4049}\\{\tiny $\pm$0.0014 $**$}} & \cellcolor{gray!15} \makecell[r]{\textbf{0.3140}\\{\tiny $\pm$0.0001 $**$}} & \cellcolor{gray!15} \makecell[r]{\textbf{0.9919}\\{\tiny $\pm$0.0000 $**$}} & \cellcolor{gray!15} \makecell[r]{\textbf{0.0288}\\{\tiny $\pm$0.0000 $**$}} & \cellcolor{gray!15} \makecell[r]{\textbf{0.0218}\\{\tiny $\pm$0.0002 $*$}} & \cellcolor{gray!15} \makecell[r]{0.0192\\{\tiny $\pm$0.0005}} & \cellcolor{gray!15} \makecell[r]{\textbf{0.9856}\\{\tiny $\pm$0.0000 $*$}} & \cellcolor{gray!15} \makecell[r]{0.0270\\{\tiny $\pm$0.0002}} \\
    \cmidrule{2-11}
     & \multirow{2}{*}{F1} & Origin & 0.4013 & 0.2206 & 0.9897 & 0.0032 & 0.1529 & \textbf{0.0876} & 0.9896 & 0.0628 \\
    & & Ours & \cellcolor{gray!15} \makecell[r]{\textbf{0.4121}\\{\tiny $\pm$0.0011 $*$}} & \cellcolor{gray!15} \makecell[r]{\textbf{0.3168}\\{\tiny $\pm$0.0001 $**$}} & \cellcolor{gray!15} \makecell[r]{\textbf{0.9959}\\{\tiny $\pm$0.0000 $**$}} & \cellcolor{gray!15} \makecell[r]{\textbf{0.0376}\\{\tiny $\pm$0.0002 $**$}} & \cellcolor{gray!15} \makecell[r]{\textbf{0.1530}\\{\tiny $\pm$0.0000 $**$}} & \cellcolor{gray!15} \makecell[r]{0.0873\\{\tiny $\pm$0.0000}} & \cellcolor{gray!15} \makecell[r]{\textbf{0.9898}\\{\tiny $\pm$0.0001 $*$}} & \cellcolor{gray!15} \makecell[r]{\textbf{0.0628}\\{\tiny $\pm$0.0000 $**$}} \\
    \midrule
    \multirow{6}{*}{DGA} & \multirow{2}{*}{AUC-PRC} & Origin & 0.2203 & 0.1175 & 0.9843 & 0.0016 & \textbf{0.1980} & \textbf{0.1315} & 0.6466 & 0.1131 \\
    & & Ours & \cellcolor{gray!15} \makecell[r]{\textbf{0.3225}\\{\tiny $\pm$0.0242 $*$}} & \cellcolor{gray!15} \makecell[r]{\textbf{0.1469}\\{\tiny $\pm$0.0014 $*$}} & \cellcolor{gray!15} \makecell[r]{\textbf{0.9891}\\{\tiny $\pm$0.0001 $**$}} & \cellcolor{gray!15} \makecell[r]{\textbf{0.0025}\\{\tiny $\pm$0.0001 $*$}} & \cellcolor{gray!15} \makecell[r]{0.1636\\{\tiny $\pm$0.0043}} & \cellcolor{gray!15} \makecell[r]{0.1269\\{\tiny $\pm$0.0058}} & \cellcolor{gray!15} \makecell[r]{\textbf{0.9827}\\{\tiny $\pm$0.0001 $**$}} & \cellcolor{gray!15} \makecell[r]{\textbf{0.1793}\\{\tiny $\pm$0.0007 $**$}} \\
    \cmidrule{2-11}
     & \multirow{2}{*}{Rec@$k$} & Origin & 0.2903 & 0.1501 & 0.9844 & 0.0000 & 0.1814 & 0.1817 & 0.7558 & 0.1652 \\
    & & Ours & \cellcolor{gray!15} \makecell[r]{\textbf{0.3990}\\{\tiny $\pm$0.0073 $*$}} & \cellcolor{gray!15} \makecell[r]{\textbf{0.2815}\\{\tiny $\pm$0.0034 $**$}} & \cellcolor{gray!15} \makecell[r]{\textbf{0.9904}\\{\tiny $\pm$0.0000 $**$}} & \cellcolor{gray!15} \makecell[r]{\textbf{0.0207}\\{\tiny $\pm$0.0000 $**$}} & \cellcolor{gray!15} \makecell[r]{\textbf{0.1951}\\{\tiny $\pm$0.0026 $*$}} & \cellcolor{gray!15} \makecell[r]{\textbf{0.1874}\\{\tiny $\pm$0.0093}} & \cellcolor{gray!15} \makecell[r]{\textbf{0.9806}\\{\tiny $\pm$0.0030 $**$}} & \cellcolor{gray!15} \makecell[r]{\textbf{0.3407}\\{\tiny $\pm$0.0022 $**$}} \\
    \cmidrule{2-11}
     & \multirow{2}{*}{F1} & Origin & 0.3795 & 0.2146 & 0.9911 & 0.0032 & 0.1534 & 0.1131 & 0.8910 & 0.0696 \\
    & & Ours & \cellcolor{gray!15} \makecell[r]{\textbf{0.4040}\\{\tiny $\pm$0.0046 $*$}} & \cellcolor{gray!15} \makecell[r]{\textbf{0.3040}\\{\tiny $\pm$0.0018 $**$}} & \cellcolor{gray!15} \makecell[r]{\textbf{0.9922}\\{\tiny $\pm$0.0003 $*$}} & \cellcolor{gray!15} \makecell[r]{\textbf{0.0224}\\{\tiny $\pm$0.0004 $**$}} & \cellcolor{gray!15} \makecell[r]{\textbf{0.3094}\\{\tiny $\pm$0.0006 $**$}} & \cellcolor{gray!15} \makecell[r]{\textbf{0.2170}\\{\tiny $\pm$0.0093 $*$}} & \cellcolor{gray!15} \makecell[r]{\textbf{0.9823}\\{\tiny $\pm$0.0020 $**$}} & \cellcolor{gray!15} \makecell[r]{\textbf{0.3414}\\{\tiny $\pm$0.0014 $**$}} \\
    \midrule
    \multirow{6}{*}{SGNN} & \multirow{2}{*}{AUC-PRC} & Origin & 0.4219 & 0.3523 & 0.9891 & \textbf{0.0039} & - & - & - & - \\
    & & Ours & \cellcolor{gray!15} \makecell[r]{\textbf{0.5584}\\{\tiny $\pm$0.0001 $**$}} & \cellcolor{gray!15} \makecell[r]{\textbf{0.4329}\\{\tiny $\pm$0.0002 $**$}} & \cellcolor{gray!15} \makecell[r]{\textbf{0.9980}\\{\tiny $\pm$0.0000 $**$}} & \cellcolor{gray!15} - & \cellcolor{gray!15} - & \cellcolor{gray!15} - & \cellcolor{gray!15} - & \cellcolor{gray!15} - \\
    \cmidrule{2-11}
     & \multirow{2}{*}{Rec@$k$} & Origin & 0.4965 & 0.3842 & 0.9778 & \textbf{0.0262} & - & - & - & - \\
    & & Ours & \cellcolor{gray!15} \makecell[r]{\textbf{0.5334}\\{\tiny $\pm$0.0002 $**$}} & \cellcolor{gray!15} \makecell[r]{\textbf{0.4600}\\{\tiny $\pm$0.0001 $**$}} & \cellcolor{gray!15} \makecell[r]{\textbf{0.9970}\\{\tiny $\pm$0.0000 $**$}} & \cellcolor{gray!15} - & \cellcolor{gray!15} - & \cellcolor{gray!15} - & \cellcolor{gray!15} - & \cellcolor{gray!15} - \\
    \cmidrule{2-11}
     & \multirow{2}{*}{F1} & Origin & 0.4959 & 0.3651 & 0.9936 & \textbf{0.0085} & - & - & - & - \\
    & & Ours & \cellcolor{gray!15} \makecell[r]{\textbf{0.5362}\\{\tiny $\pm$0.0002 $**$}} & \cellcolor{gray!15} \makecell[r]{\textbf{0.4718}\\{\tiny $\pm$0.0000 $**$}} & \cellcolor{gray!15} \makecell[r]{\textbf{0.9982}\\{\tiny $\pm$0.0000 $**$}} & \cellcolor{gray!15} - & \cellcolor{gray!15} - & \cellcolor{gray!15} - & \cellcolor{gray!15} - & \cellcolor{gray!15} - \\
    \bottomrule
\end{tabular}
    }
    \caption{Overall comparison results on the remaining transfer settings. $**$ indicates $p<0.001$ and $*$ indicates $p<0.1$ in the paired t-test compared with the baseline.}
    \label{tab:tta_comp_2}
\end{table*}

Tables~\ref{tab:tta_comp_1} and~\ref{tab:tta_comp_2} compare the performance of ``Original'' models without TTA and ``Ours'' with \NAME. 
Here, ``-'' denotes out-of-memory (OOM) errors, ``tr.'' and ``te.'' denote ``train on'' and ``test on'', respectively.
Overall, \NAME~ preserves the anomaly detection ability learned from the source dataset while adapting the model to shifted test distributions.

Following Table~\ref{tab:dataset}, we use the anomaly ratio of each test dataset as the random AUC-PRC baseline. 
When the original model performs poorly, such as transferring from CryptopiaHacker to PlusTokenPonzi, the source and target datasets exhibit clear distribution shifts, and \NAME~ consistently improves performance through test-time adaptation.
When the original model already performs well, such as transferring from AlphaHomora to CryptopiaHacker, \NAME~ still maintains or further improves the performance, indicating that the adaptation process does not severely damage source-domain knowledge.

The OOM cases mainly occur when combining the large Trace dataset with SpaceGNN. 
During TTA, maintaining both teacher and student models further increases memory consumption, making SpaceGNN infeasible on Trace under our hardware setup.

\subsection{Case Study on Temporal Motif Patterns}

\begin{figure}[t]
    \centering
    \includegraphics[width=1\linewidth]{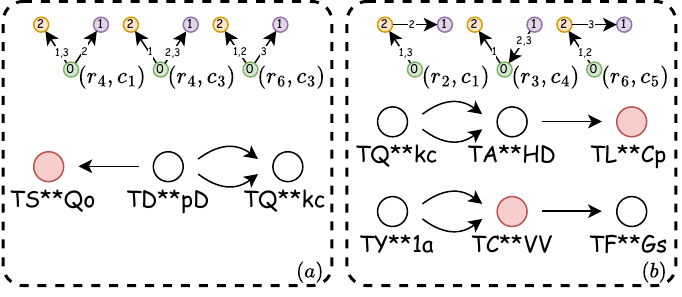}
    \caption{Common temporal motif patterns in real-world blockchain datasets.}
    \label{fig:case_study}
\end{figure}

To verify whether temporal motifs can capture meaningful anomalous behaviors, we analyze common motif patterns in our private dataset. 
As shown in Figure~\ref{fig:case_study}, we group motifs by their basic structures and present representative cases verified by our partner.

Group (a) shows a distribution pattern from node-0. 
In our cases, the most common receiver in this pattern is a private exchange service frequently used in fraud cases, while the other receiver is the suspicious node labeled as fraud.
Although such distribution patterns are common on blockchains, they become strong fraud signals within a short time window because malicious actors often avoid using the same account for a long period.

Group (b) shows aggregation patterns, where assets are collected into one account before payment. 
Due to temporal-order constraints, only motif $(r_6,c_3)$ clearly indicates this behavior.
In the first case, the middle node aggregates assets from an exchange service and sends them to the fraud-related target node.
In the second case, the aggregation node itself is suspicious, corresponding to a cash-out pattern according to our partner.
This indicates that fraud-related addresses may repeatedly receive assets from upstream addresses and forward them downstream, forming distinctive temporal aggregation motifs.

\subsection{Ablation Study}

\begin{table}[t]
    \centering
    \scalebox{0.80}{
        \begin{tabular}{ll|rrrr}
    \toprule
    Dataset & Metric & GCN & SAGE & DGA & SGNN \\
                    \midrule
    \multirow{4}{*}{Alpha} & AUC-PRC & \cellcolor{gray!15} \textbf{0.7173} & \cellcolor{gray!15} \textbf{0.7368} & \cellcolor{gray!15} \textbf{0.6756} & \cellcolor{gray!15} 0.7713 \\
                      & \it{w/o motif} & 0.6034 & 0.7214 & 0.6323 & \textbf{0.7753} \\
                      & Rec@$k$ & \cellcolor{gray!15} \textbf{0.6676} & \cellcolor{gray!15} \textbf{0.6849} & \cellcolor{gray!15} \textbf{0.6224} & \cellcolor{gray!15} 0.7159 \\
                      & \it{w/o motif} & 0.5653 & 0.6758 & 0.5839 & \textbf{0.7179} \\
    \midrule
    \multirow{4}{*}{Crypto} & AUC-PRC & \cellcolor{gray!15} \textbf{0.5713} & \cellcolor{gray!15} 0.4611 & \cellcolor{gray!15} \textbf{0.4768} & \cellcolor{gray!15} 0.7051 \\
                      & \it{w/o motif} & 0.4869 & \textbf{0.5223} & 0.4404 & \textbf{0.7061} \\
                      & Rec@$k$ & \cellcolor{gray!15} \textbf{0.5655} & \cellcolor{gray!15} 0.5158 & \cellcolor{gray!15} \textbf{0.4526} & \cellcolor{gray!15} 0.6588 \\
                      & \it{w/o motif} & 0.4745 & \textbf{0.5164} & 0.4474 & \textbf{0.6614} \\
    \midrule
    \multirow{4}{*}{Plus} & AUC-PRC & \cellcolor{gray!15} 0.9960 & \cellcolor{gray!15} 0.9919 & \cellcolor{gray!15} 0.9931 & \cellcolor{gray!15} 0.9987 \\
                      & \it{w/o motif} & \textbf{0.9968} & \textbf{0.9944} & \textbf{0.9955} & \textbf{0.9993} \\
                      & Rec@$k$ & \cellcolor{gray!15} \textbf{0.9908} & \cellcolor{gray!15} 0.9736 & \cellcolor{gray!15} 0.9798 & \cellcolor{gray!15} 0.9966 \\
                      & \it{w/o motif} & 0.9869 & \textbf{0.9902} & \textbf{0.9881} & \textbf{0.9968} \\
    \midrule
    \multirow{4}{*}{Upbit} & AUC-PRC & \cellcolor{gray!15} \textbf{0.4346} & \cellcolor{gray!15} 0.4114 & \cellcolor{gray!15} \textbf{0.3553} & \cellcolor{gray!15} \textbf{0.7329} \\
                      & \it{w/o motif} & 0.4023 & \textbf{0.5221} & 0.3480 & 0.7177 \\
                      & Rec@$k$ & \cellcolor{gray!15} \textbf{0.4788} & \cellcolor{gray!15} 0.4482 & \cellcolor{gray!15} 0.4100 & \cellcolor{gray!15} \textbf{0.7173} \\
                      & \it{w/o motif} & 0.4783 & \textbf{0.5357} & \textbf{0.4528} & 0.7063 \\
    \bottomrule
\end{tabular}
    }
    \caption{Ablation study on GNNs with motif representation.}
    \label{tab:motifs_performance}
\end{table}

\paragraph{Temporal Motif.}
To verify the effectiveness of the proposed motif representation, we conduct ablation experiments on four public datasets with different GNN backbones.
As shown in Table~\ref{tab:motifs_performance}, motif representation consistently improves anomaly detection performance, especially on AlphaHomora and UpbitHack.
GCN obtains the most significant improvement, suggesting that motif features can compensate for its limited structural expressiveness.
The minor improvement on PlusTokenPonzi may be due to its high anomaly ratio and nearly saturated prediction performance, leaving limited room for further gains.

\begin{table*}[!htbp]
    \centering
    \scalebox{0.88}{
        \begin{tabular}{l|l|l|rrrr|rrrr|rrrr}
    \toprule
        \multicolumn{3}{c}{Model} & \multicolumn{4}{|c}{GCN} & \multicolumn{4}{|c}{SAGE} & \multicolumn{4}{|c}{SGNN} \\
    \midrule
        dataset & $k$ & ~$\Delta t$~ & None & 1800 & 3600 & 7200 & None & 1800 & 3600 & 7200 & None & 1800 & 3600 & 7200  \\
            \midrule
        \multirow{4}{*}{Alpha} & \multicolumn{2}{l|}{None}  & - & 0.6077 & 0.7172 & 0.7117& - & 0.5876 & 0.7310 & 0.7213& - & 0.7512 & 0.7677 & 0.7699 \\
                           & \multicolumn{2}{l|}{50}    & 0.4357 & 0.6406 & \underline{0.7195} & 0.7144& 0.3928 & 0.6606 & 0.7254 & 0.7209& 0.7347 & 0.7215 & \textbf{0.7742} & 0.7720 \\
                           & \multicolumn{2}{l|}{\cellcolor{gray!15}100}   & \cellcolor{gray!15} 0.3952 & \cellcolor{gray!15} 0.6556 & \cellcolor{gray!15} \textbf{0.7205} & \cellcolor{gray!15} 0.7145& \cellcolor{gray!15} 0.2397 & \cellcolor{gray!15} 0.6503 & \cellcolor{gray!15} \underline{0.7327} & \cellcolor{gray!15} 0.7268& \cellcolor{gray!15} 0.7373 & \cellcolor{gray!15} 0.7494 & \cellcolor{gray!15} \underline{0.7727} & \cellcolor{gray!15} 0.7695 \\
                           & \multicolumn{2}{l|}{200}   & 0.3549 & 0.6448 & 0.7111 & 0.7160& 0.2801 & 0.6006 & \textbf{0.7338} & 0.7282& 0.7312 & 0.7520 & 0.7675 & 0.7673 \\        \midrule
        \multirow{4}{*}{Crypto} & \multicolumn{2}{l|}{None}  & - & 0.4214 & \underline{0.5774} & \textbf{0.5855}& - & 0.3050 & 0.5268 & 0.4944& - & 0.7013 & 0.6945 & 0.6844 \\
                           & \multicolumn{2}{l|}{50}    & 0.3136 & 0.5264 & 0.5276 & 0.5610& 0.1302 & 0.4030 & 0.4903 & \textbf{0.5666}& 0.6734 & 0.6860 & 0.6840 & 0.6812 \\
                           & \multicolumn{2}{l|}{\cellcolor{gray!15}100}   & \cellcolor{gray!15} 0.3118 & \cellcolor{gray!15} 0.3962 & \cellcolor{gray!15} 0.5749 & \cellcolor{gray!15} 0.5594& \cellcolor{gray!15} 0.1303 & \cellcolor{gray!15} 0.2982 & \cellcolor{gray!15} 0.4830 & \cellcolor{gray!15} 0.4931& \cellcolor{gray!15} 0.6824 & \cellcolor{gray!15} 0.6853 & \cellcolor{gray!15} 0.6986 & \cellcolor{gray!15} \underline{0.7033} \\
                           & \multicolumn{2}{l|}{200}   & 0.2625 & 0.3664 & 0.5473 & 0.5663& 0.1012 & 0.2064 & 0.5205 & \underline{0.5504}& 0.6725 & \textbf{0.7105} & 0.6969 & 0.6856 \\        \midrule
        \multirow{4}{*}{Plus} & \multicolumn{2}{l|}{None}  & - & 0.9972 & 0.9970 & 0.9974& - & 0.9909 & \textbf{0.9952} & 0.9925& - & 0.9983 & \textbf{0.9988} & 0.9985 \\
                           & \multicolumn{2}{l|}{50}    & 0.9957 & \underline{0.9975} & 0.9964 & 0.9962& 0.9930 & 0.9932 & 0.9914 & 0.9930& \underline{0.9987} & 0.9981 & 0.9986 & \underline{0.9987} \\
                           & \multicolumn{2}{l|}{\cellcolor{gray!15}100}   & \cellcolor{gray!15} 0.9957 & \cellcolor{gray!15} 0.9966 & \cellcolor{gray!15} 0.9964 & \cellcolor{gray!15} 0.9957& \cellcolor{gray!15} 0.9938 & \cellcolor{gray!15} 0.9926 & \cellcolor{gray!15} 0.9920 & \cellcolor{gray!15} 0.9934& \cellcolor{gray!15} \underline{0.9987} & \cellcolor{gray!15} 0.9984 & \cellcolor{gray!15} \underline{0.9987} & \cellcolor{gray!15} 0.9985 \\
                           & \multicolumn{2}{l|}{200}   & 0.9953 & \textbf{0.9977} & 0.9973 & 0.9969& 0.9932 & 0.9934 & 0.9944 & \underline{0.9949}& \underline{0.9987} & 0.9983 & \underline{0.9987} & 0.9985 \\        \midrule
        \multirow{4}{*}{Upbit} & \multicolumn{2}{l|}{None}  & - & 0.2074 & 0.3719 & 0.3946& - & 0.3636 & 0.3288 & 0.3317& - & \textbf{0.7418} & 0.7406 & 0.7349 \\
                           & \multicolumn{2}{l|}{50}    & 0.1676 & 0.3889 & \textbf{0.4376} & \underline{0.4294}& 0.1016 & 0.4120 & \underline{0.4408} & \textbf{0.4783}& 0.7347 & 0.7379 & 0.7294 & 0.7322 \\
                           & \multicolumn{2}{l|}{\cellcolor{gray!15}100}   & \cellcolor{gray!15} 0.1553 & \cellcolor{gray!15} 0.3690 & \cellcolor{gray!15} 0.4045 & \cellcolor{gray!15} 0.4132& \cellcolor{gray!15} 0.0407 & \cellcolor{gray!15} 0.4156 & \cellcolor{gray!15} 0.4390 & \cellcolor{gray!15} 0.4398& \cellcolor{gray!15} 0.7317 & \cellcolor{gray!15} 0.7337 & \cellcolor{gray!15} 0.7345 & \cellcolor{gray!15} 0.7375 \\
                           & \multicolumn{2}{l|}{200}   & 0.1297 & 0.3170 & 0.3972 & 0.4128& 0.0308 & 0.3069 & 0.3248 & 0.3483& 0.7324 & 0.7364 & 0.7343 & \underline{0.7410} \\
    \bottomrule
\end{tabular}
    }
    \caption{Parameter analysis results of motif matching in AUC-PRC.}
    \label{tab:param_motifs}
\end{table*}

\begin{table}[!htbp]
    \centering
    \scalebox{1}{
        \begin{tabular}{lrrr}
    \toprule
    Dataset & w/ $k$ (s) & w/ $\Delta t$ (s) & w/ $k, \Delta t$ (s) \\
    \midrule
        Alpha & 1,307.01 & \textbf{174.06} & 176.04 \\
        Crypto & 2,134.01 & \textbf{129.03} & 139.08 \\
        Plus & 20.08 & 21.02 & \textbf{19.02} \\
        Upbit & 462.07 & 63.10 & \textbf{61.08} \\
        \rowcolor{gray!15} 
        Trace & 732.01 & 3,005.07 & \textbf{349.05} \\
    \bottomrule
\end{tabular}
    }
    \caption{Running time of motif matching.}
    \label{tab:motifs_time}
\end{table}

\begin{table}[!htbp]
    \centering
    \scalebox{0.9}{
            \begin{tabular}{l|rrr|rrr}
        \toprule
        \multirow{2}{*}{$\alpha$} & \multicolumn{3}{c}{Alpha} & \multicolumn{3}{c}{Crypto} \\
        \cmidrule{2-7}
         & Crypto & Plus & Upbit & Crypto & Plus & Upbit \\
        \midrule
        \rowcolor{gray!15} 
        0.9 & 0.3850 & \textbf{0.8501} & \textbf{0.2564} & \textbf{0.6253} & \textbf{0.9713} & 0.4108\\
        0.99 & 0.3889 & 0.8283 & 0.2672 & 0.6241 & 0.9712 & 0.4300\\
        0.999 & \textbf{0.3892} & 0.8242 & 0.2678 & 0.6240 & \textbf{0.9713} & \textbf{0.4312} \\
        \bottomrule
    \end{tabular}
    }
    \caption{AUC-PRC results of the momentum coefficient $\alpha$.}
    \label{tab:param_alpha}
\end{table}

\begin{table}[!htbp]
    \centering
    \scalebox{0.8}{
            \begin{tabular}{ll|rrr|rrr}
        \toprule
        \multirow{2}{*}{$\tau_{low}$} & \multirow{2}{*}{$\tau_{high}$} & \multicolumn{3}{c}{Alpha} & \multicolumn{3}{c}{Crypto} \\
        \cmidrule{3-8}
         & & Crypto & Plus & Upbit & Crypto & Plus & Upbit \\
        \midrule
        \multirow{3}{*}{0.5} & 0.8 & \textbf{0.3309} & 0.9471 & 0.3030 & 0.6359 & 0.8290 & 0.2983\\
         & 0.9 & 0.3295 & 0.9509 & 0.3061 & \textbf{0.6367} & \textbf{0.8308} & 0.2978\\
         & 0.95 & 0.3281 & 0.9469 & \textbf{0.3123} & 0.6366 & 0.8291 & 0.2981\\
        \midrule
        \multirow{3}{*}{0.7} & 0.8 & 0.3270 & \textbf{0.9616} & 0.3066 & 0.6360 & 0.8288 & 0.2972\\
         & 0.9 & 0.3288 & 0.9438 & 0.3034 & 0.6357 & 0.8245 & 0.2977\\
         & 0.95 & 0.3257 & 0.9222 & 0.3058 & 0.6349 & 0.8305 & 0.2979\\
        \midrule
        \multirow{1}{*}{0.9} & 0.95 & 0.3255 & 0.9499 & 0.2969 & 0.6363 & 0.8285 & \textbf{0.2997} \\
        \bottomrule
    \end{tabular}
    }
    \caption{AUC-PRC results of the trustable node threshold $\tau$.}
    \label{tab:param_tau}
\end{table}

\subsection{Parameter Analysis}

\paragraph{Edge limit $k$ and time aggregation range $\Delta t$.} 
We analyze the impact of $k$ and $\Delta t$ on motif representation, with results shown in Table~\ref{tab:param_motifs}.
In most settings, a proper edge limit and time aggregation range improve anomaly detection performance.
The edge limit focuses motif extraction on temporally relevant transactions, while time aggregation reduces duplicate transaction patterns.
We further evaluate the runtime efficiency of motif matching on real-world transaction graphs, as reported in Table~\ref{tab:motifs_time}.
The results show that $k$ and $\Delta t$ substantially reduce computational cost, demonstrating the practical scalability of our motif extraction method on large-scale blockchain datasets with over 1 million nodes and 2 million edges.

In deployment, \NAME~ runs on localized $k$-hop subgraphs around seed accounts instead of the full blockchain graph. 
Motif matching costs about 0.12 ms per edge, and forward/backward passes take 40.25/169.69 ms, leading to second-level latency under Ethereum-scale throughput and supporting real-time on-chain fraud detection.

\paragraph{Teacher model momentum $\alpha$.}
The teacher momentum $\alpha$ controls how much information the teacher model receives from the student model.
We evaluate different values of $\alpha$ using GCN.
As shown in Table~\ref{tab:param_alpha}, a larger $\alpha$ usually weakens TTA performance, such as in the transfer from AlphaHomora to PlusTokenPonzi, because it slows down knowledge adaptation from the student model.

\paragraph{Trustable node selection thresholds $\tau_{low}$ and $\tau_{high}$.} 
The thresholds $\tau_{low}$ and $\tau_{high}$ define the confidence range for selecting trustable nodes, as summarized in Table~\ref{tab:param_tau}.
Increasing $\tau_{high}$ or decreasing $\tau_{low}$ generally expands the selected node set and helps the model capture more useful adaptation signals.
However, this benefit only holds within a proper range: an excessively high $\tau_{high}$ may introduce over-confident predictions, while an overly low $\tau_{low}$ may include noisy low-confidence nodes, both of which degrade TTA performance.

\section{Conclusion}

In this work, we propose \NAME, a temporal motif-aware graph test-time adaptation framework for blockchain anomaly detection. 
\NAME~ integrates 3-node temporal motif representations with original node features to capture fine-grained higher-order transaction patterns. 
It further introduces a simple yet effective test-time adaptation strategy to mitigate distribution shifts caused by evolving malicious behaviors and adversarial transaction patterns.

Extensive experiments on real-world blockchain datasets demonstrate that \NAME~ consistently outperforms classical and \emph{state-of-the-art} graph anomaly detection models. 
The case study on temporal motif patterns provides interpretable evidence that our method can characterize meaningful fraud-related transaction structures. 
Ablation studies further verify the effectiveness of both temporal motif representation and test-time adaptation, showing how different components contribute to the overall performance.

Beyond experimental evaluation, \NAME~ has also been validated in collaboration with a public security department. 
In 2025, it was applied to real-world blockchain transaction data and supported the detection of suspicious transactions involving more than 1 million USDT. 
This practical validation demonstrates the potential of \NAME~ for real-time on-chain risk monitoring and provides evidence of its real-world impact in blockchain fraud detection.

For future work, it is promising to incorporate large language models (LLMs) into blockchain anomaly detection, given their strong zero-shot inference and natural-language understanding capabilities. 
LLMs could help integrate heterogeneous information sources, such as smart-contract code, transaction metadata, and off-chain reports, thereby improving detection performance while providing more interpretable explanations for detected anomalies.

\FloatBarrier

\section*{Acknowledgments}

This work is supported by the Zhejiang Province ``JianBingLingYan+X'' Research and Development Plan (2025C02020).



\bibliographystyle{named}
\bibliography{main}

\end{document}